# FIRST PRINCIPLES INVESTIGATION OF THE STRAIN-MODE COUPLING IN SrBi$_2$Nb$_2$O$_9$

Urko Petralanda and I. Etxebarria

Fisika Aplikatua II Saila, Zientzia eta Teknologia, Euskal Herriko Unibertsitatea,
P. K. 644 48080 Bilbao, Spain
inigo.etxebarria@ehu.es

**Abstract**
The Aurivillius compounds SrBi$_2$Nb$_2$O$_9$ (SBN) and SrBi$_2$Ta$_2$O$_9$ (SBT) present at room temperature isomorphous ferroelectric structures that are stabilized by a complex interplay of three order parameters via a trilinear coupling. In this work we examine the influence of the in-plane misfit strain in the strength of the relevant instabilities and the couplings between them. A mixed strain-stress enthalpy compatible with the symmetry of SBN is developed with all the parameters obtained from first principles. We find that the dominant role of the order parameters varies as a function of the strain and, as a consequence, that the value of the polarization can be tuned by choosing the lattice parameter of the substrate. It is also shown that the sequence of phase transitions and the main features of the phase diagram change qualitatively in terms of the strain.

**Introduction**
The Aurivillius compounds SrBi$_2$Nb$_2$O$_9$ (SBN) and SrBi$_2$Ta$_2$O$_9$ (SBT) are ferroelectric at room temperatures, and due to their low fatigue degradation they have been considered as potential candidates for nonvolatile memories and thoroughly investigated experimentally [1-3] and computationally [4-7]. The polar low-temperature structure ($A2_1am$) can be described as a distortion with respect to the tetragonal parent phase ($I4/mmm$) where three symmetry adapted modes of $X_3^-$, $\Gamma_5^-$ and $X_2^+$ symmetry are frozen. The $X_3^-$ mode lowers the tetragonal symmetry to the orthorhombic *Amam* (No. 63) space group, and consist of rotations of the oxygen octahedra as rigid units about the tetragonal $(1,1,0)_T$ direction and displacements of the Bi cations along the $(1,-1,0)_T$ direction. The polar $\Gamma_5^-$ distortion involves essentially an antiphase displacement of the Bi atoms and the perovskite blocks along $(1,1,0)_T$, and its condensation produces a ferroelectric structure with *F2mm* (No. 46) space group. Finally, the $X_2^+$ mode is compatible with the *Abam* (No. 64) space group and corresponds to tiltings of the oxygen octahedra about the $z$ axis and antiphase displacements of the oxygens in the Bi$_2$O$_2$ slabs along the $(1,1,0)_T$ direction. On increasing the temperature SBN presents a single first order phase transition $A2_1am \rightarrow I4/mmm$, while in the case of SBT an intermediate phase of symmetry *Amam* is observed.

The simultaneous presence of at least two of the three distortions yields the symmetry of the ground state. According to their dominant amplitudes, the $X_3^-$ and $\Gamma_5^-$ modes are the natural candidates. However, it has been shown [5-6] that the presence of the $X_2^+$ mode is essential to stabilize the ferroelectric phase via a trilinear coupling. A similar mechanism has been found in the Ruddlesden-Popper Ca$_2$Mn$_2$O$_7$, double perovskites as NaLaMnWO$_6$ or Sr$_2$MWO$_6$ (M=Zn, Ca, Mg), and even in the simple perovskite PbZrO$_3$ [8-12]

The properties of perovskite thin films are strongly influenced by the epitaxial strain resulting from lattice-matching the film to the substrate [13-16]. The studies of the coupling between strain and symmetry modes are mainly focused on compounds where the instabilities are governed by only one zone center order parameter. In this work we study the effect of strain on the phase diagram and critical instabilities of SBN, extending the formalism to a system with three order parameters.

**The model and methodology**
The energy per unit cell can be divided into three terms: the pure strain energy, the energy associated to the pure modes and the interaction between the strain and the modes.

The pure elastic energy of the tetragonal *I4/mmm* structure up to second order is:

$$E_{\text{elastic}} = \frac{1}{2}B_{11}(\eta_1^2 + \eta_2^2) + \frac{1}{2}B_{33}\eta_3^2 + B_{12}\eta_1\eta_2 + B_{13}(\eta_1 + \eta_2)\eta_3 + \frac{1}{2}B_{44}(\eta_4^2 + \eta_5^2) + \frac{1}{2}B_{66}\eta_6^2 \quad (2)$$

where the Voigt notation has been used.
The contribution from the three modes up to fourth order is given by [5-6]:

$$E_{\text{mode}} = \frac{1}{2}\kappa_{X_3^-}Q_{X_3^-}^2 + \beta_{X_3^-}Q_{X_3^-}^4 + \frac{1}{2}\kappa_{\Gamma_5^-}Q_{\Gamma_5^-}^2 + \beta_{\Gamma_5^-}Q_{\Gamma_5^-}^4 + \frac{1}{2}\kappa_{X_2^+}Q_{X_2^+}^2 + \beta_{X_2^+}Q_{X_2^+}^4 + \\ \delta_{X_3^-\Gamma_5^-}Q_{X_3^-}^2 Q_{\Gamma_5^-}^2 + \delta_{X_3^-X_2^+}Q_{X_3^-}^2 Q_{X_2^+}^2 + \delta_{\Gamma_5^-X_2^+}Q_{\Gamma_5^-}^2 Q_{X_2^+}^2 + \gamma Q_{X_3^-}Q_{\Gamma_5^-}Q_{X_2^+} \quad (1)$$

And the strain/mode interaction reads up to first order in strain and third order in the mode amplitudes reads:

$$E_{\text{inter}} = \frac{1}{2}B_{1X_3^-}(\eta_1 + \eta_2)Q_{X_3^-}^2 + \frac{1}{2}B_{3X_3^-}\eta_3 Q_{X_3^-}^2 + \frac{1}{2}B_{6X_3^-}\eta_6 Q_{X_3^-}^2 + \\ \frac{1}{2}B_{1\Gamma_5^-}(\eta_1 + \eta_2)Q_{\Gamma_5^-}^2 + \frac{1}{2}B_{3\Gamma_5^-}\eta_3 Q_{\Gamma_5^-}^2 + \frac{1}{2}B_{6\Gamma_5^-}\eta_6 Q_{\Gamma_5^-}^2 + \\ \frac{1}{2}B_{1X_2^+}(\eta_1 + \eta_2)Q_{X_2^+}^2 + \frac{1}{2}B_{3X_2^+}\eta_3 Q_{X_2^+}^2 + \frac{1}{2}B_{6X_2^+}\eta_6 Q_{X_2^+}^2 + \\ B_{1X_3^-\Gamma_5^-X_2^+}(\eta_1 + \eta_2)Q_{X_3^-}Q_{\Gamma_5^-}Q_{X_2^+} + B_{3X_3^-\Gamma_5^-X_2^+}\eta_3 Q_{X_3^-}Q_{\Gamma_5^-}Q_{X_2^+} + B_{6X_3^-\Gamma_5^-X_2^+}\eta_6 Q_{X_3^-}Q_{\Gamma_5^-}Q_{X_2^+} \quad (3)$$

where the coupling between the shear strains $\eta_4$ and $\eta_5$ and the three modes are forbidden by symmetry in this order.

In epitaxy, the strain components $\eta_1$, $\eta_2$ and $\eta_6$ are constrained, and the most convenient mixed stress-strain enthalpy can be constructed from $G^* = E - \sigma_3\eta_3 - \sigma_4\eta_4 - \sigma_5\eta_5$ with the natural variables $\eta_1$, $\eta_2$, $\eta_6$, $\sigma_3$, $\sigma_4$, $\sigma_5$, $Q_{X_3^-}$, $Q_{\Gamma_5^-}$ and $Q_{X_2^+}$ [17-18]. Minimizing $G^*$ with respect to $\eta_3$, $\eta_4$ and $\eta_5$, and assuming that the shear stresses $\sigma_4$ and $\sigma_5$, and the perpendicular $\sigma_3$ stress vanish we obtain:

$$\eta_3 = -\frac{1}{B_{33}}\left[\frac{1}{2}B_{3X_3^-}Q_{X_3^-}^2 + \frac{1}{2}B_{3\Gamma_5^-}Q_{\Gamma_5^-}^2 + \frac{1}{2}B_{3X_2^+}Q_{X_2^+}^2 + B_{3X_3^-\Gamma_5^-X_2^+}Q_{X_3^-}Q_{\Gamma_5^-}Q_{X_2^+}\right]$$

$$\eta_4 = \eta_5 = 0 \quad (4)$$

We are mainly concerned with the effects of epitaxial strain on a substrate with square symmetry. In consequence, the tetragonal symmetry is conserved: $\eta_1 = \eta_2 = \bar{\eta}$ and $\eta_6 = 0$. $\bar{\eta} = (a-a_0)/a$ is the misfit strain between the square substrate ($a$) and the $a_0 = b_0$ lattice parameters of SBN. Substituting this constraints and Eq. (4) back to Eq. (3), the enthalpy can be expressed in terms of $\bar{\eta}$, $Q_{X_3^-}$, $Q_{\Gamma_5^-}$ and $Q_{X_2^+}$:

$$G^* = B_{\bar{\eta}}\bar{\eta}^2 + \frac{1}{2}\kappa^*_{X_3^-}Q_{X_3^-}^2 + \beta^*_{X_3^-}Q_{X_3^-}^4 + \frac{1}{2}\kappa^*_{\Gamma_5^-}Q_{\Gamma_5^-}^2 + \beta^*_{\Gamma_5^-}Q_{\Gamma_5^-}^4 + \frac{1}{2}\kappa^*_{X_2^+}Q_{X_2^+}^2 + \beta^*_{X_2^+}Q_{X_2^+}^4 +$$
$$\delta^*_{X_3^-\Gamma_5^-}Q_{X_3^-}^2Q_{\Gamma_5^-}^2 + \delta^*_{X_3^-X_2^+}Q_{X_3^-}^2Q_{X_2^+}^2 + \delta^*_{\Gamma_5^-X_2^+}Q_{\Gamma_5^-}^2Q_{X_2^+}^2 + \gamma^*Q_{X_3^-}Q_{\Gamma_5^-}Q_{X_2^+} +$$
$$\mu_{X_3^-}Q_{X_3^-}^3Q_{\Gamma_5^-}Q_{X_2^+} + \mu_{\Gamma_5^-}Q_{X_3^-}Q_{\Gamma_5^-}^3Q_{X_2^+} + \mu_{X_2^+}Q_{X_3^-}Q_{\Gamma_5^-}Q_{X_2^+}^3 + \omega Q_{X_3^-}^2Q_{\Gamma_5^-}^2Q_{X_2^+}^2 \quad (5)$$

where the renormalized coefficients are:

$$B_{\bar{\eta}} = B_{11} + B_{12}$$

$$\kappa^*_{X_3^-} = \kappa_{X_3^-} + 2B_{\bar{\eta}X_3^-}\bar{\eta} \qquad \kappa^*_{\Gamma_5^-} = \kappa_{\Gamma_5^-} + 2B_{\bar{\eta}\Gamma_5^-}\bar{\eta} \qquad \kappa^*_{X_2^+} = \kappa_{X_2^+} + 2B_{\bar{\eta}X_2^+}\bar{\eta}$$

$$\beta^*_{X_3^-} = \beta_{X_3^-} - \frac{B_{3X_3^-}^2}{8B_{33}} \qquad \beta^*_{\Gamma_5^-} = \beta_{\Gamma_5^-} - \frac{B_{3\Gamma_5^-}^2}{8B_{33}} \qquad \beta^*_{X_2^+} = \beta_{X_2^+} - \frac{B_{3X_2^+}^2}{8B_{33}}$$

$$\delta^*_{X_3^-\Gamma_5^-} = \delta_{X_3^-\Gamma_5^-} - \frac{B_{3X_3^-}B_{3\Gamma_5^-}}{4B_{33}} \qquad \delta^*_{X_3^-X_2^+} = \delta_{X_3^-X_2^+} - \frac{B_{3X_3^-}B_{3X_2^+}}{4B_{33}} \qquad \delta^*_{\Gamma_5^-X_2^+} = \delta_{\Gamma_5^-X_2^+} - \frac{B_{3\Gamma_5^-}B_{3X_2^+}}{4B_{33}}$$

$$\mu_{X_3^-} = -\frac{B_{3X_3^-}B_{3X_3^-\Gamma_5^-X_2^+}}{2B_{33}} \qquad \mu_{\Gamma_5^-} = -\frac{B_{3\Gamma_5^-}B_{3X_3^-\Gamma_5^-X_2^+}}{2B_{33}} \qquad \mu_{X_2^+} = -\frac{B_{3X_2^+}B_{3X_3^-\Gamma_5^-X_2^+}}{2B_{33}}$$

$$\gamma^* = \gamma + 2B_{\bar{\eta}X_3^-\Gamma_5^-X_2^+}\bar{\eta} \qquad \omega = -\frac{B_{3X_3^-\Gamma_5^-X_2^+}^2}{2B_{33}}$$

$$B_{\bar{\eta}X_3^-} = B_{1X_3^-} - \frac{B_{3X_3^-}B_{13}}{2B_{33}} \qquad B_{\bar{\eta}\Gamma_5^-} = B_{1\Gamma_5^-} - \frac{B_{3\Gamma_5^-}B_{13}}{2B_{33}} \qquad B_{\bar{\eta}X_2^+} = B_{1X_2^+} - \frac{B_{3X_2^+}B_{13}}{2B_{33}}$$

$$B_{\bar{\eta}X_3^-\Gamma_5^-X_2^+} = B_{1X_3^-\Gamma_5^-X_2^+} - \frac{B_{3X_3^-\Gamma_5^-X_2^+}B_{13}}{2B_{33}}$$

Several new high order terms arise in the expansion of the enthalpy due to the renormalization of the $B_{3X_3^-\Gamma_5^-X_2^+}$ coupling. Among them, the sixth order coefficient ($\omega$) is negative and the enthalpy should be corrected by including also higher order terms in Eq. (2) (10 new coefficients up to sixth order). However, we chose to limit the number of parameters of the model by neglecting the tiny $B_{3X_3^-\Gamma_5^-X_2^+}$ coupling and its effect in the renormalized enthalpy ($\omega = \mu_{X_3^-} = \mu_{\Gamma_5^-} = \mu_{X_2^+} = 0$).

We used the full potential LAPW+lo method as implemented in WIEN2k [19]. Exchange and correlation effects were treated within the GGA approximation with the Perdew-Burke-Ernzehof (PBE) parametrization [20]. The $RK_{max}$ parameter related to the number of radial basis functions taken to describe the inside spheres was chosen to be 7.0. Calculations were performed using a Monkhorst-Pack $k$ point mesh of $4 \times 4 \times 4$, equivalent to 12 independent $k$ points in the irreducible Brillouin wedge. The radii of the atomic spheres were 2.26 (Sr), 2.26 (Bi), 1.80 (Nb) and 1.46 (O) bohrs. Energy convergence tests to determine the accuracy of the parameters preceded the calculations. First, relaxation of ground state was carried out obtaining forces below 0.1 mRy/bohr. The convergence criterion for the SCF calculations was of 0.1 mRy for energy and 0.1 mRy/bohr for forces. First, the system was relaxed at zero strain giving a structure and symmetry mode decomposition similar to that of Ref. [6]. Then, the *ab initio* energies at more than 900 derived structures were calculated in the range $-0.2 < \bar{\eta}, \eta_3 < 0.2$ and $0 < Q_{X_3^-}, Q_{\Gamma_5^-}, Q_{X_2^+} < 3.5$ bohr of the configuration space. Finally, these energies were used to determine the coefficients of the enthalpy expansion by least squares fitting.

**Energetics and instabilities**
The energy of the ferroelectric ground state and of each pure mode was analized in terms of the misfit strain. Fig. 1 shows the depth of the energy wells associated to the pure distortions ($X_3^-, \Gamma_5^-$ and $X_2^+$) and to the absolute minimum with respect to the tetragonal structure. The evolution of the renormalized trilinear coupling ($\gamma^*$) in mRy/bohr$^3$ is also shown. Although in the region of large tensile strain $\gamma^*$ is close to zero, the ferroelectric $A2_1am$ structure, where the three modes are condensed, is the absolute minimum for all the values of $\bar{\eta}$ studied. However, the influence of strain is essential to assign a dominant role between the three order parameters. For zero misfit strain the energy gains of $X_3^-$ and $\Gamma_5^-$ are quite similar while the slight instability of the $X_2^+$ distortion shows its secondary role. In the region of tensile stress ($\bar{\eta} > 0$) the $X_2^+$ mode becomes stable, and the depths associated to $\Gamma_5^-$ and $X_3^-$ grow; for large positive misfit strains the polar instability is dominant and the energy gain of the $A2_1am$ phase with respect to the ferroelectric structure ($F2mm$) associated to the condensation of $\Gamma_5^-$ alone is very small. The scenario is the opposite for compressive strains: $X_3^-$ and $\Gamma_5^-$ become stable and the "secondary" $X_2^+$ mode is dominant (*Abam*). Similar conclusions can be obtained from Fig 2., where the contributions of the three distortions to the structure of the ground state are depicted. The amplitudes of $\Gamma_5^-$ and $X_2^+$ are higher for tensile and compressive strains respectively, and the amplitude of the $X_3^-$ mode shows a weaker dependence on the misfit strain. The change of sign in the amplitude in the $X_2^+$ distortion occurs at the value where the renormalized trilinear coupling vanishes (Fig. 1). These results indicate that the epitaxial grown of SBN on a square substrate with a larger lattice parameter could be used to enhance the polarization associated to the $\Gamma_5^-$ ferroelectric instability.

**Phase diagram**

The expansion of the energy in the previous section can be considered as the zero temperature free energy of the system, and a phenomenological free energy can be approximated under the assumptions of the Landau theory of phase transitions. Then, the temperature renormalization is solely determined by the quadratic terms and the stiffness constants present a linear dependence on temperature, such that $\kappa_i^* = a_i(T - T_{o,i})$ for a given $i$ mode, where $a_i$ and $T_{o,i}$ are constants related to the *ab initio* stiffness constants by $\kappa_i^* = \kappa_i^*(0) = -a_i T_{o,i}$. For instance, the temperature evolution of the quadratic term in the enthalpy for the $X_3^-$ mode is:

$$\kappa_{X_3^-}^*(T) = a_{X_3^-}(T - T_{o,X_3^-}) = \kappa_{X_3^-}^*(0) + a_{X_3^-} T = \kappa_{X_3^-} + 2 B_{\bar{\eta} X_3^-} \bar{\eta} + a_{X_3^-} T$$

Given the similar qualitative dependence of $X_3^-$ and $\Gamma_5^-$ with respect to the strain (Fig. 1), we have calculated several sections of the phase diagrams that correspond to a homogeneous temperature renormalization of both order parameters, that is, the difference between $\kappa_{X_3^-}^*(T)$ and $\kappa_{\Gamma_5^-}^*(T)$ remains constant along each diagram according to $\kappa_{\Gamma_5^-}^*(T) - \kappa_{X_3^-}^*(T) = \kappa_{\Gamma_5^-}^*(0) - \kappa_{X_3^-}^*(0) = \Delta$ with $\Delta = +2.776765, +0.4494$ and $-1.87796$ for $\bar{\eta} = -0.05, +0.0$ and $+0.05$ respectively. Fig. 3 shows sections of the phase diagram for three representative values of the in-plane strain. At zero strain there is a wide passage that joins the stability regions of the tetragonal and ferroelectric phases. Thus, an appropriate renormalization of the order parameters could drive the system directly from the ferroelectric state to the parent structure without an intermediate phase as observed in the *avalanche* phase transition of SBN. The presence of a negative strain ($\bar{\eta} = -0.05$) narrows the direct passage although it does not change qualitatively the main features of the diagram and the limits of the stability regions remain similar. However, at zero temperature the stiffness constants of $X_3^-$ and $\Gamma_5^-$ are close to zero and the effect of a homogeneous temperature renormalization could change the sequence of phase transitions to $A2_1am \rightarrow Abam \rightarrow I4/mmm$. Finally, the effect of a positive misfit strain ($\bar{\eta} = +0.05$) produces qualitative changes, the topology of the phase diagram forbids a direct transition between the high- and low-temperature structures. Moreover, the narrow stability region of the *Amam* phase associated to the condensation of the $X_3^-$ order parameter is replaced by a wide *F2mm* region, that correspond to a polar structure with only the $\Gamma_5^-$ order parameter frozen. This result suggests that under a positive in-plane strain the sequence of phase transitions in terms of temperature could change to $A2_1am \rightarrow F2mmm \rightarrow I4/mmm$ with an intermediate ferroelectric phase.

**Conclusions**

The development of a mixed strain-stress enthalpy compatible with the tetragonal high temperature of SBN, the symmetry of three order-parameters involved and the application of an in-plane strain reveals qualitative changes in the behaviour of the system.

The ground state of SBN for all the values of misfit strain studied does not change

($A2_1am$). However, the role of the relevant modes varies significantly: at negative strain the $X_2^+$ mode is dominant whereas at positive strains the polar $\Gamma_5^-$ mode is the main instability. This feature could be used to tune the polarization of SBN by choosing the lattice parameter of the substrate.

Regarding the phase diagram, results show that the sequence of phase transitions when increasing the temperature could be different for positive and negative strains. For $\bar{\eta} = -0.05$ an intermediate phase of *Abam* symmetry should be present whereas a ferroelectric *F2mmm* structure should appear for $\bar{\eta} = +0.05$.

All the calculations have been done by fixing the orientation of the order parameters along the direction that is active in the ground state [6]. Thus, the two-dimensional character of the $X_3^-, \Gamma_5^-$ and $X_2^+$ irreducible representations has been eluded by working effectively with three one-dimensional order parameters. The consequences of this approximation cannot be easily evaluated because an exhaustive description would result in a dramatic increment in the number of parameters to be obtained.

**Figure captions**

**Figure 1:** Depths of the energy wells per formula unit associated to the pure distortions $X_3^-$ (open circles), $\Gamma_5^-$ (open squares) and $X_2^+$ (solid squares) and to the absolute minimum (solid circles) with respect to the relaxed tetragonal parent configuration. The evolution of the renormalized trilinear coupling ($\gamma^*$) in mRy/bohr$^3$ is also shown.

**Figure 2:** Mode decomposition of the absolute minimum ($A2_1am$) of Fig. 1 with respect to the tetragonal parent phase. The amplitudes are given for modes normalized within the primitive unit cell of the ferroelectric structure.

**Figure 3:** Phase diagram of SBN as a function of the $\kappa^*_{X_3^-}(T)$ and $\kappa^*_{X_2^+}(T)$ stiffness constants for three values of the misfit strain. The renormalization of the polar mode $\kappa^*_{\Gamma_5^-}(T)$ is implicitly taken into account as explained in the text.

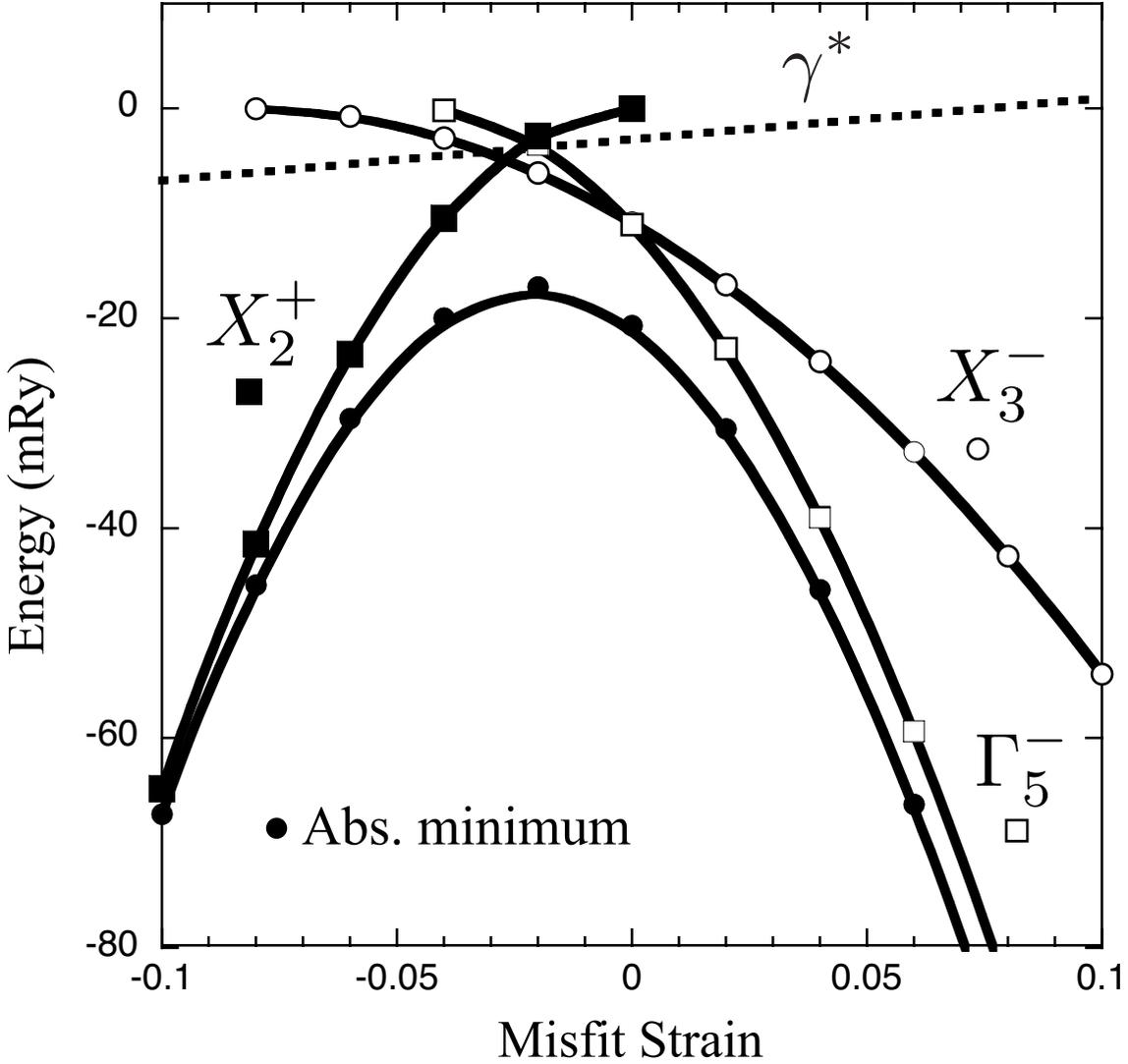

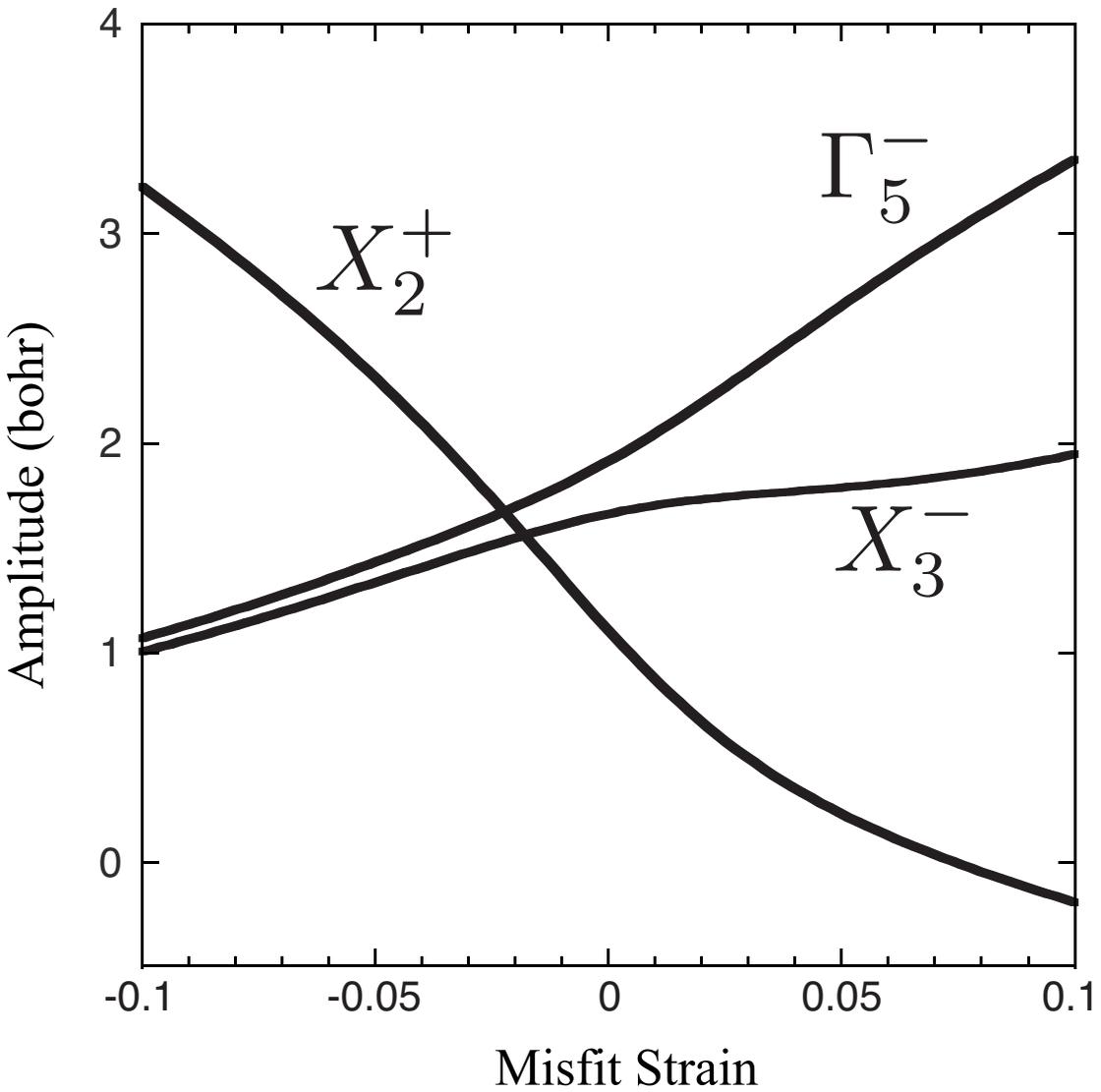

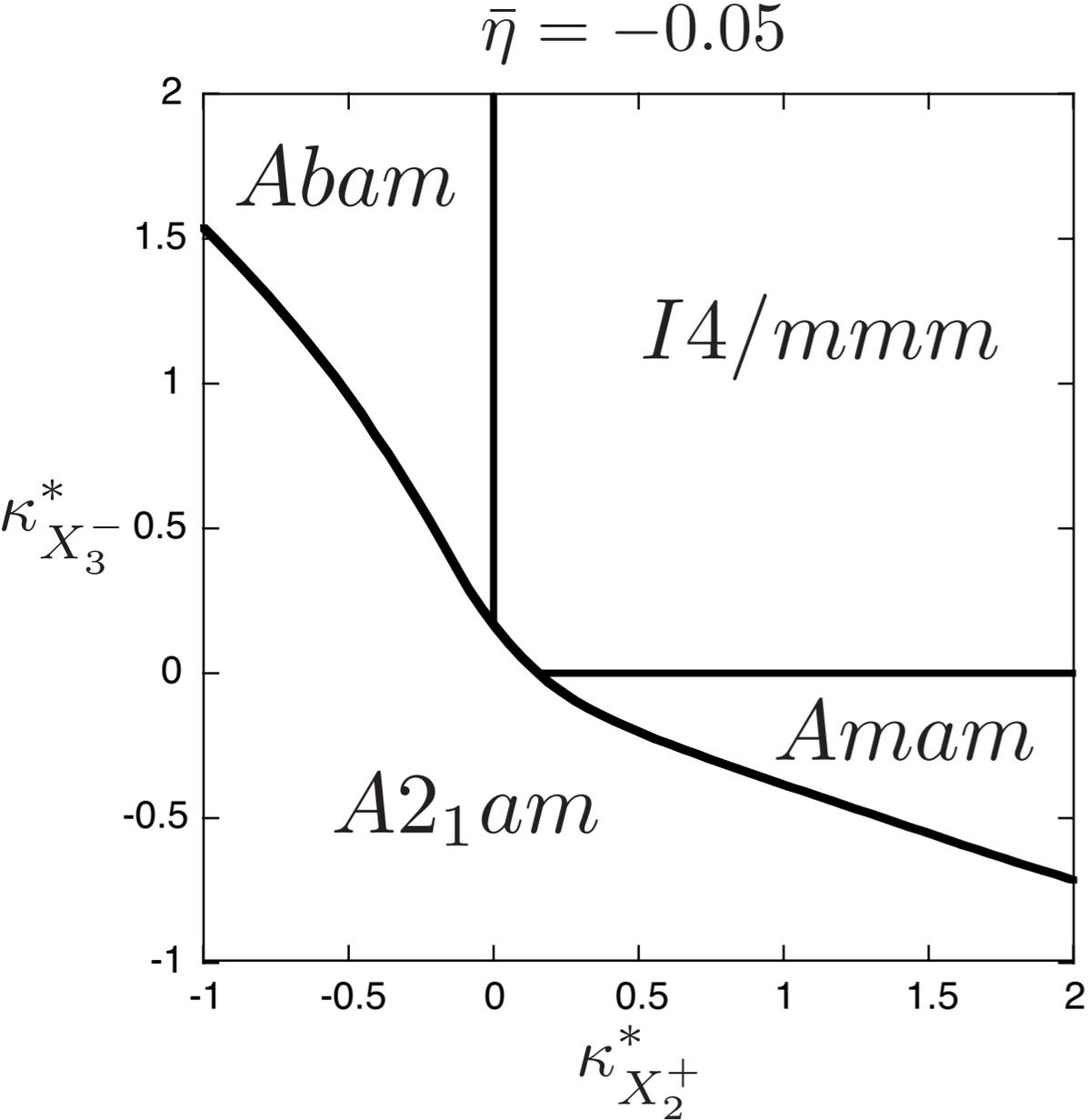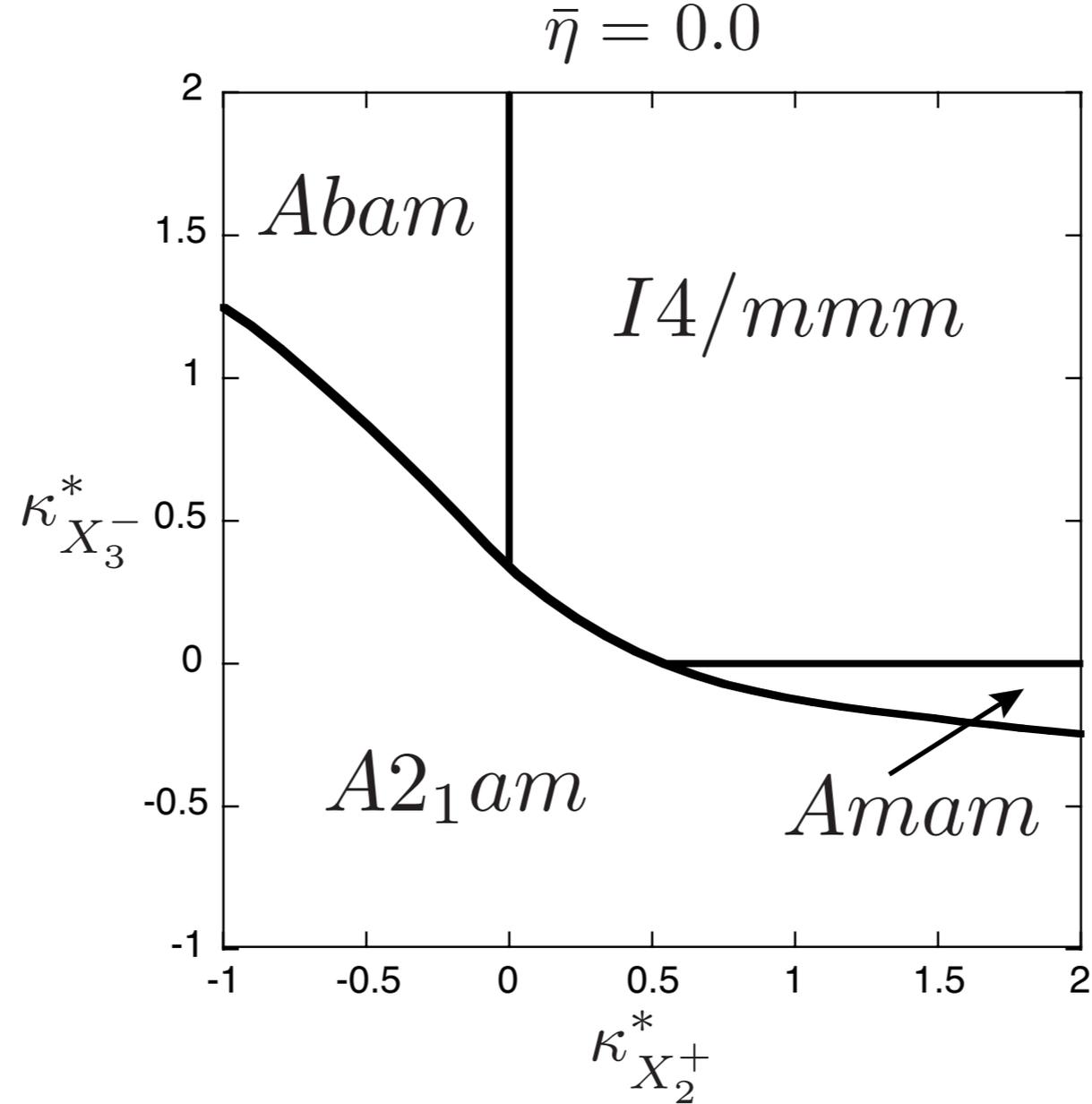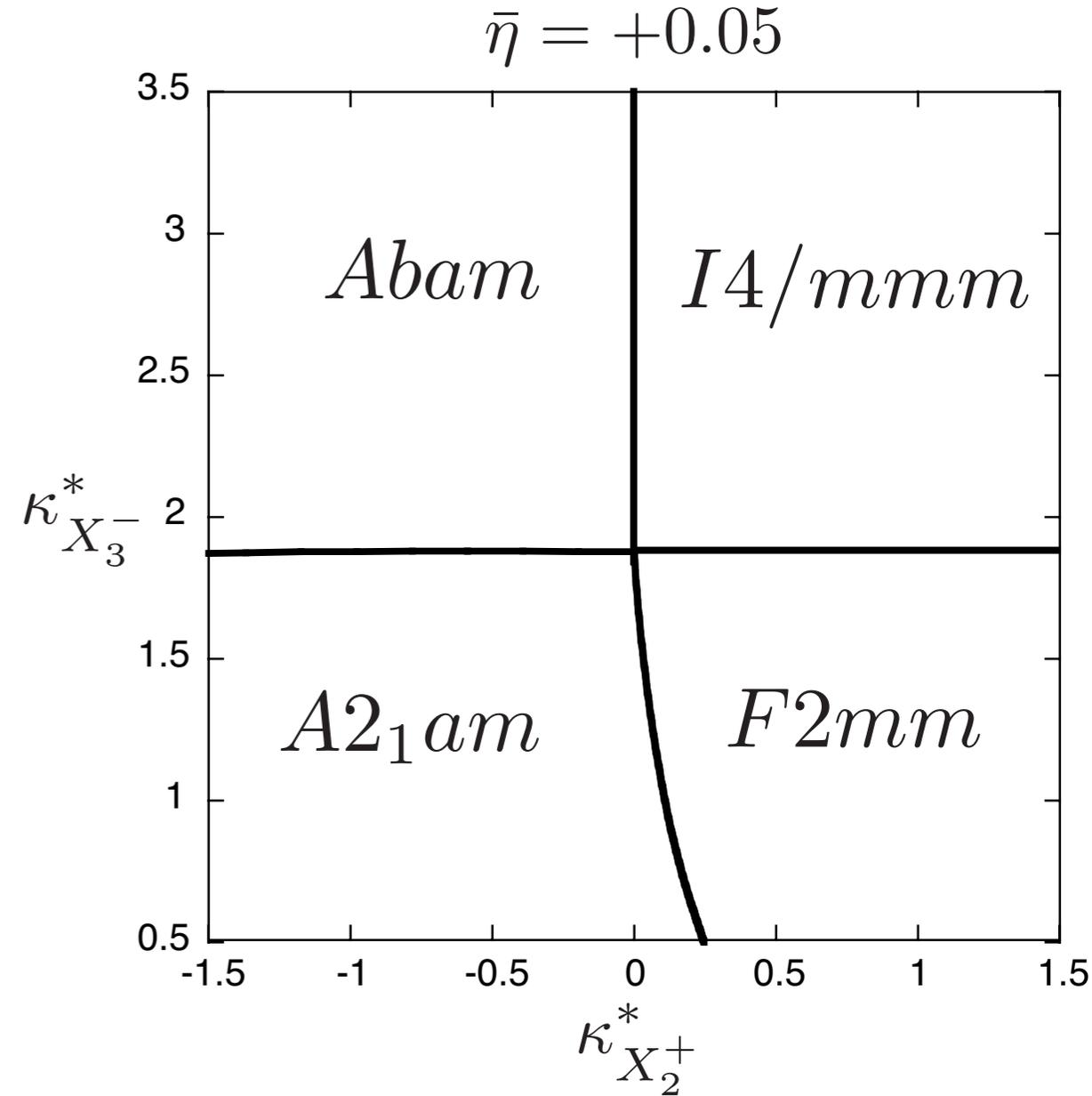